\documentclass[10pt, conference, compsocconf]{IEEEtran}

\usepackage{algorithm,algorithmic,amsbsy,amsmath,amssymb,epsfig,subfig,bbm,mathrsfs,mathdots,arydshln,arydshln,multirow,verbatim,booktabs,setspace,url,stfloats,fixltx2e,cite,nomencl,epstopdf}

\usepackage{braket}

\usepackage{xcolor}
\def\BibTeX{{\rm B\kern-.05em{\sc i\kern-.025em b}\kern-.08em
    T\kern-.1667em\lower.7ex\hbox{E}\kern-.125emX}}

\newcommand{\beq}{\begin{equation}}
\newcommand{\eeq}{\end{equation}}
\newcommand{\bitm}{\begin{itemize}}
\newcommand{\ba}{\begin{array}}
\newcommand{\ea}{\end{array}}
\newcommand{\eitm}{\end{itemize}}
\newcommand{\beqn}{\begin{eqnarray}}
\newcommand{\eeqn}{\end{eqnarray}}
\newcommand{\beqno}{\begin{eqnarray*}}
\newcommand{\eeqno}{\end{eqnarray*}}
\newcommand{\bma}{\begin{displaymath}}
\newcommand{\ema}{\end{displaymath}}
\newcommand{\bnu}{\begin{enumerate}}
\newcommand{\enu}{\end{enumerate}}
\newcommand{\bce}{\begin{center}}
\newcommand{\ece}{\end{center}}
\newcommand{\btb}{\begin{tabular}}
\newcommand{\etb}{\end{tabular}}

\begin{document}

\title{Stochastic Qubit Resource Allocation for Quantum Cloud Computing}
\author{\IEEEauthorblockN{Rakpong Kaewpuang$^{\mathrm{1}}$, Minrui Xu$^{\mathrm{1}}$, Dusit Niyato$^{\mathrm{1}}$, Han Yu$^{\mathrm{1}}$, Zehui Xiong$^{\mathrm{2}}$, and Jiawen Kang$^{\mathrm{3}}$} \\
\IEEEauthorblockA{ $^{\mathrm{1}}$School of Computer Science and Engineering, Nanyang Technological University, Singapore \\
				    $^{\mathrm{2}}$Pillar of Information Systems Technology and Design, Singapore University of Technology and Design, Singapore \\
                    $^{\mathrm{3}}$School of Automation, Guangdong University of Technology, China
       }}
\maketitle

\begin{abstract}
Quantum cloud computing is a promising paradigm for efficiently provisioning quantum resources (i.e., qubits) to users. In quantum cloud computing, quantum cloud providers provision quantum resources in reservation and on-demand plans for users. Literally, the cost of quantum resources in the reservation plan is expected to be cheaper than the cost of quantum resources in the on-demand plan. However, quantum resources in the reservation plan have to be reserved in advance without information about the requirement of quantum circuits beforehand, and consequently, the resources are insufficient, i.e., under-reservation. Hence, quantum resources in the on-demand plan can be used to compensate for the unsatisfied quantum resources required. To end this, we propose a quantum resource allocation for the quantum cloud computing system in which quantum resources and the minimum waiting time of quantum circuits are jointly optimized. Particularly, the objective is to minimize the total costs of quantum circuits under uncertainties regarding qubit requirement and minimum waiting time of quantum circuits. In experiments, practical circuits of quantum Fourier transform are applied to evaluate the proposed qubit resource allocation. The results illustrate that the proposed qubit resource allocation can achieve the optimal total costs.                    
\end{abstract}


\section{Introduction and Related Works}    
\label{sec:introduction}
Quantum cloud computing \cite{quantum-ai-google2022} opens up a new potential to address large-scale simulation and optimization problems in communication and network systems. Through superposition, entanglement, and interference of quantum bits, i.e., qubits, quantum cloud computing can outperform classical cloud computing and existing supercomputers by speeding up computations and reducing energy consumption. In the Noisy Intermediate-Scale Quantum (NISQ) era \cite{k-bharti-noisy2022}, \cite{j-preskill-quantum-computing-nisq2018}, quantum cloud computing offered by AWS~\cite{amazon-barket2022}, IBM~\cite{ibm-quantum-computing}, and Azure~\cite{azure-quantum2022} is beginning to show its capabilities by transforming the existing fields of machine learning, security, and finance.

In quantum cloud computing, quantum resources such as qubits are scarcer and more expensive than in traditional cloud computing~\cite{s-resch-benchmarking-quantum2022}. In particular, the performance of quantum computing is affected not only by the number of qubits but also by the depth of the quantum circuit and the noise at each point of the quantum circuit~\cite{a-wack-qss2021}. First, the scale of quantum cloud computing determines the size of the quantum computing task that can be encoded and solved. Second, the quality indicates the size of quantum circuits that can be implemented and executed for quantum computing tasks and algorithms. Finally, the speed of quantum cloud computing depends on the number of quantum circuits and the number of executions that a quantum computer can perform per unit of time.

Similar to traditional cloud computing~\cite{s-chaisiri-ovmp2009}, a quantum cloud user must acquire the required quantum computing resources from the quantum cloud service provider through a subscription. During execution, quantum cloud operators specify the computation demand and parameters to be provided in terms of qubits, depending on the difficulty of each user's computing task. Users can acquire quantum computing resources from the quantum cloud operator through both reservation and on-demand plans. Specifically, the quantum cloud user reserves the quantum computing resources from the quantum cloud operator according to an estimated task difficulty and waiting time expected by the user. During the execution of the quantum computing task, the quantum cloud user also purchases additional quantum computing resources from the operator in case of insufficient reserved quantum computing resources under uncertainties of qubit requirement. In addition, the quantum cloud user has to wait until the quantum circuit (i.e., computing task) is successfully executed. 

In related research works, the resource allocation models in distributed quantum computing were presented in~\cite{n-ngoenriang-optimal-stochastic2022},~\cite{g-s-ravi-quantum-computing-in-cloud2021},~\cite{g-s-ravi-adaptive-job2021}, and~\cite{c-cicconetti-resource-allocation-quantum2022}. However, the existing works do not consider the problem of jointly optimizing the qubit resource allocation and waiting time for quantum circuits in quantum cloud computing. In addition, current works omit the uncertainties of the number of qubits and the waiting time of quantum circuits that affect the total costs of quantum circuits.

The major contributions of this paper can be summarized as follows:
\begin{itemize}
 \item We propose a quantum resource allocation for the quantum cloud computing system for jointly optimizing the cost of quantum circuits under the uncertainty of quantum applications and waiting time. Specifically, a two-stage stochastic programming (SP) model is proposed to minimize the total costs of quantum circuits under uncertainties of qubit requirements and waiting time of quantum circuits.
 \item We show the preeminent performance of the proposed model by experimenting with the circuit requirement of quantum Fourier transform (QFT) under practical quantum computing programming environments
\end{itemize}

\section{System Model and Assumptions}

\begin{figure}[htb]
\begin{center}
$\begin{array}{c} \epsfxsize=2.5 in \epsffile{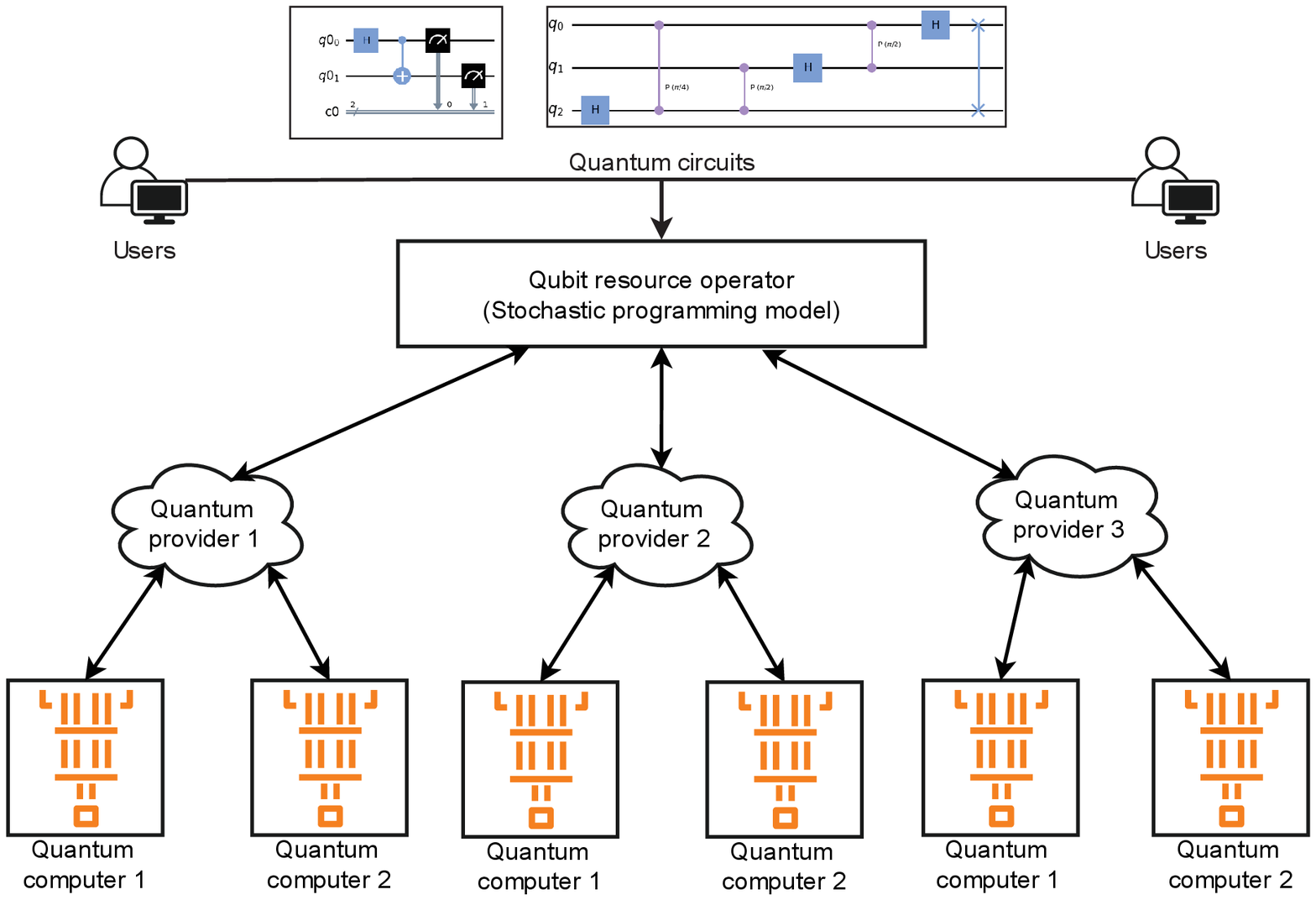} \\
\end{array}$
\caption{Quantum cloud computing environment.} 
\label{fig:system-model}
\end{center}
\vspace{-0.4cm}
\end{figure}

We consider the system model of quantum cloud computing illustrated in Fig.~\ref{fig:system-model}. We consider that the system model comprises users, a qubit resource operator, quantum providers, and quantum computers. The users possess quantum circuits that will be executed on quantum computers of quantum providers. A set of quantum circuits is denoted as $\mathcal{C}$. Along with the quantum circuits submitted to quantum providers, the users request a certain number of qubits and specify the waiting time for the quantum circuits that will be successfully completed. The quantum providers provide users with quantum computing resources (i.e., qubits in quantum computers). The set of quantum providers is denoted as $\mathcal{P}$. Quantum computers, which can support quantum circuits execution, with different numbers of qubits are located in quantum data centers and managed by quantum providers. The set of quantum computers in quantum provider $p$ is denoted as $\mathcal{M}_{p}$.  

In this paper, we consider the number of qubits required by users and the waiting time of the quantum circuits as uncertain demand and requirement, respectively. In quantum computing resource provisioning, the quantum providers offer the users two subscription plans that consist of the reservation plan and the on-demand plan, similar to the payment plans of conventional cloud computing \cite{s-chaisiri-optimization2011}, \cite{s-chaisiri-ovmp2009}. The cost of the reservation plan is lower than that of the on-demand plan. Let $R^{\mathrm{qp}}_{c,p}$, $U^{\mathrm{qp}}_{c,p}$, and $O^{\mathrm{qp}}_{c,p}$ denote a reservation cost of qubit, a utilized cost of qubit, an on-demand cost of qubit for quantum circuit $c$ charged by provider $p$, respectively. 

The quantum providers will charge the users when quantum computing resources are reserved or utilized. We consider that there are three phases of operations \cite{s-chaisiri-optimization2011}, \cite{s-chaisiri-ovmp2009}: reservation, utilization, and on-demand for provisioning quantum computing resources. In the reservation phase, quantum providers provision quantum computing resources without information about users' demands. Afterward, the utilization phase is activated when the computing resources reserved in the reservation phase are used to execute the quantum circuits. Nevertheless, if the computing resources reserved in the reservation phase cannot satisfy all the demands, the quantum providers can provide the computing resources in the on-demand phase to fulfill the unfulfilled demands. In addition, the users need the quantum circuits to be executed successfully and as fast as possible. Therefore, we consider that users can specify the waiting time for the quantum circuits (i.e., $\alpha_{c,\omega}$) to be successfully executed. The cost of over-waiting time will be charged, i.e., as a penalty, to quantum providers if the quantum circuits execution cannot be finished by the time specified by users. Let $P^{\mathrm{pt}}_{c,p}$ denote a penalty cost of over-waiting time of quantum circuit $c$ from provider $p$ and $t^{\mathrm{exe}}_{c,p,m}$ denote a real execution time of quantum circuit $c$ executed on quantum computer $m$ of provider $p$. 

 When the execution time of circuit $c$ executed on machine $m$ of provider $p$ is higher than the waiting time of circuit $c$ specified by the user (i.e., $t^{\mathrm{exe}}_{c,p,m} > \alpha_{c,\omega}$), the over-waiting time (i.e., $y_{c,p,m,\omega} > 0$) occurs and the cost of over-waiting time is charged to provider $p$. Otherwise, there is no over-waiting time of circuit $c$ (i.e., $y_{c,p,m,\omega} = 0$) since quantum machine $m$ executes quantum circuit $c$ on time. 

Therefore, quantum providers face a stochastic optimization problem to minimize the total cost (i.e., the cost associated with provisioning quantum computing resources and the penalty cost of over-waiting time) while satisfying uncertainties of users' demands and waiting time for quantum circuits. The qubit resource operator is introduced to provision the number of quantum computing resources to support quantum circuits and to reduce the over-waiting time for quantum circuits. The qubit resource operator is developed based on the two-stage SP.

\section{Qubit Resource Allocation Formulation}
\label{qubit-resource-allocation-formulation}
  
\subsection{Model Description}
\label{subsec:model-description}

The optimization is based on the two-stage SP. The decision variables in the optimization are defined as follows: 
\begin{itemize}
	\item $x^{\mathrm{r}}_{c,p,m}$ denotes a non-negative integer variable representing the number of qubits that are reserved for quantum circuit $c$ on quantum computer $m$ of provider $p$ in the reservation phase.
	\item $x^{\mathrm{u}}_{c,p,m,\omega}$ denotes a non-negative integer variable representing the number of qubits that are utilized by quantum circuit $c$ on quantum computer $m$ of provider $p$.  
	\item  $x^{\mathrm{o}}_{c,p,\omega}$ denotes a non-negative integer variable representing the number of on-demand qubits that are utilized by quantum circuit $c$ in provider $p$.
	\item  $y_{c,p,m,\omega}$ denotes a positive real variable representing the over-waiting time of quantum circuit $c$ running on quantum computer $m$ of provider $p$. 
\end{itemize}

We consider uncertainties of the number of qubits required by quantum circuits and the waiting time for quantum circuits to be completed. Let $\mathcal{B}_{c}$ and $\mathcal{E}_{c}$ respectively denote the set of the possible number of qubits required by quantum circuit $c$ from a user and the set of execution time for the quantum circuit $c$, which are expressed as follows: 
\beqn
 \mathcal{B}_{c} = \{ \beta_{c1}, \beta_{c2}, \dots, \beta_{c{\varphi}}\}, \mathcal{E}_{c} = \{ \alpha_{c1}, \alpha_{c2}, \dots, \alpha_{c\varpi}\}. \label{set_eq} 
\eeqn
Where $\varphi$ and $\varpi$ are the last indexes of elements in the finite sets $\mathcal{B}_{c}$ and $\mathcal{E}_{c}$, respectively. Let $\tilde{\omega}$ denote the composite random variable expressed as follows: 
\beqn
\tilde{\omega} \in \Big\{ \big( \tilde{\beta}_{c}, \tilde{\alpha}_{c} \big) \Big| \label{omega_eq}  \tilde{\beta}_{c} \in \mathcal{B}_{c}, \tilde{\alpha}_{c} \in \mathcal{E}_{c} \Big\}. 
\eeqn
Where $\tilde{\beta}_{c}$ and $\tilde{\alpha}_{c}$ denote the random variables of the number of qubits required by circuit $c$ from a user and the waiting time for quantum circuit $c$, respectively.
 
\subsection{Optimization Formulation}
\label{subsec:def}

The SP with the random variable $\tilde{\omega}$ can be transformed into the deterministic equivalence problem~\cite{Brige1997}. The composite random variable $\tilde{\omega}$ can be described by a \emph{scenario}. Let $\omega$ denote the scenario that is the realization of a random variable. The value of the random variable can be taken from a set of scenarios. Let $\Omega$ denote the set of all scenarios (i.e., a scenario space) in the second stage. $\Omega$ is defined as $\Omega = \mathcal{B}_{c} \times \mathcal{E}_{c}$, where $\times$ is the Cartesian product. $\mathcal{B}_{c}$ and $\mathcal{E}_{c}$ as defined in Eq.~(\ref{set_eq}) are the sets of the possible number of qubits and waiting time for quantum circuit $c$, respectively. The expectation of the SP model is defined as the weighted sum given the probability of each scenario in $\mathbb{P}(\omega)$.

\begin{figure}[htb]
\beqn
   & & \min_{x^{\mathrm{r}}_{c,p,m}, x^{\mathrm{o}}_{c,p,m,\omega}, y_{c,p,m,\omega} } \sum_{c \in \mathcal{C}} \sum_{p \in \mathcal{P}} \sum_{m \in \mathcal{M}_{p}} x^{\mathrm{r}}_{c,p,m} R^{\mathrm{qp}}_{c,p} + \nonumber \\
   & & \sum_{c \in \mathcal{C}} \sum_{p \in \mathcal{P}} \sum_{m \in \mathcal{M}_{p}} \sum_{\omega \in \Omega} \mathbb{P}(\omega) ( x^{\mathrm{u}}_{c,p,m,\omega} U^{\mathrm{qp}}_{c,p} + \nonumber \\
   & & \quad\quad\quad\quad\quad\quad\quad x^{\mathrm{o}}_{c,p,m,\omega} O^{\mathrm{qp}}_{c,p} + y_{c,p,m,\omega} P^{\mathrm{pt}}_{c,p}) \label{def-obj}   \\ 
	 \mbox{s.t.} & & x^{\mathrm{u}}_{c,p,m,\omega} \leq  x^{\mathrm{r}}_{c,p,m}, \label{def-con1} \nonumber \\
        & & \forall c \in \mathcal{C}, \forall p \in \mathcal{P}, \forall m \in \mathcal{M}_{p}, \forall \omega \in \Omega, \\    
	    & & x^{\mathrm{u}}_{c,p,m,\omega} + x^{\mathrm{o}}_{c,p,m,\omega} \geq  \beta_{c,\omega}, \label{def-con2} \nonumber \\
        & & \forall c \in \mathcal{C}, \forall p \in \mathcal{P}, \forall m \in \mathcal{M}_{p}, \forall \omega \in \Omega,\\	
        & & t^{\mathrm{exe}}_{c,p,m} \leq  \alpha_{c,\omega} +  y_{c,p,m,\omega},   \label{def-con3} \nonumber \\
        & & \forall c \in \mathcal{C}, \forall p \in \mathcal{P}, \forall m \in \mathcal{M}_{p}, \forall \omega \in \Omega, \\	
        & & x^{\mathrm{r}}_{c,p,m} \leq Q^{\mathrm{qbt}}_{p,m}, \forall c \in \mathcal{C}, \forall p \in \mathcal{P}, \forall m \in \mathcal{M}_{p}, \label{def-con4}  \\   
	    & & x^{\mathrm{r}}_{c,p,m}, x^{\mathrm{u}}_{c,p,m,\omega}, x^{\mathrm{o}}_{c,p,m,\omega} \in  \mathbb{Z}^{\ast} \; {\mathrm{and}} \; y_{c,p,m,\omega} \in \mathbb{R}^{+}, \nonumber \\
        & & \forall c \in \mathcal{C},  \forall p \in \mathcal{P}, \forall m \in \mathcal{M}_{p}, \forall \omega \in \Omega \label{def-con5}.
\eeqn
\end{figure}

The objective function in Eq.~(\ref{def-obj}) is to minimize the total costs of qubit utilization and penalty due to over-waiting time for executing all quantum circuits. The decision variables $x^{\mathrm{u}}_{c,p,m,\omega}$, $x^{\mathrm{o}}_{c,p,m,\omega}$, and $y_{c,p,m,\omega}$ are determined under scenario $\omega$ ($\omega \in \Omega$), which represents that the number of qubits required by quantum circuits and waiting time for quantum circuits are observed when $\omega$ is available.

The constraint in Eq.~(\ref{def-con1}) ensures that the number of qubits utilized by quantum circuits does not exceed the number of qubits reserved in the first stage. The constraint in Eq.~(\ref{def-con2}) ensures that the sum of the number of utilized qubits and on-demand qubits must satisfy the total qubit requirements of the quantum circuit (i.e., the number of qubits). The constraint in Eq.~(\ref{def-con3}) ensures that the waiting time of the quantum circuit ($\alpha_{c,\omega}$) is satisfied. If the execution time of quantum circuit $c$ executed on quantum computer $m$ of the quantum provider $p$ ($t^{\mathrm{exe}}_{c,p,m}$) is more than the waiting time of quantum circuit $c$, adding the over-waiting time of quantum circuit $c$ ($y_{c,p,m,\omega}$) will be charged. The constraint in Eq.~(\ref{def-con4}) ensures that the number of qubits reserved for quantum circuits is not more than the maximum number of qubits of the quantum providers ($Q^{\mathrm{qbt}}_{p,m}$). The constraint in Eq.~(\ref{def-con5}) ensures that all decision variables are non-negative integers except for $y_{c,p,m,\omega}$ that is a positive real decision variable. 

\section{Performance Evaluation}
\label{sec:performanceEvaluation}

\subsection{Parameter Setting}

We consider the quantum cloud computing system shown in Fig.~\ref{fig:system-model} which comprises three quantum cloud providers, each with two quantum computers. We set $Q^{\mathrm{qbt}}_{p,m} = 30$ for the maximum number of qubits of each quantum computer $m$ in provider $p$. We initially set $P^{\mathrm{pt}}_{c,p}$ = 10\$ for the penalty cost of the waiting time for the quantum circuit $c$ handled by provider $p$. For cost values of provider $p$ to be charged for executing the quantum circuit $c$, we initially set $R^{\mathrm{qp}}_{c,p}$ = 1.68\$~\cite{ibm-quantum-computing}, $U^{\mathrm{qp}}_{c,p}$ = 0.1\$, and  $O^{\mathrm{qp}}_{c,p}$ = 7\$ for the reservation cost, utilization cost, and on-demand cost, respectively. We consider the quantum circuits of QFT \cite{quantum-fourier-transform-qiskit2022}. We experiment with the quantum circuits with different numbers of qubits of QFT. The execution time of quantum circuit $c$ executed on quantum computer $m$ of provider $p$ ($t^{\mathrm{exe}}_{c,p,m}$) is measured from the Qiskit simulation \cite{quantum-fourier-transform-qiskit2022}. We implement and solve the SP model by using GAMS/CPLEX solver~\cite{Gams}.  

For the SP model, the random number of qubits required by the quantum circuit is ranged between 10 and 22 qubits (i.e., $\beta_{c,\omega} \in \{10,\dots,22\}$) while the random waiting time of the quantum circuit is ranged between 0.001 and 0.009 seconds (i.e., $\alpha_{c,\omega} \in \{0.001,\dots,0.009\}$). We consider both random variables with uniform distribution for ease of presentation.

\subsection{Numerical Results}

\subsubsection{Quantum Circuits and Execution Times of QFT}

We perform experiments with the QFT's quantum circuits implemented by Qiskit~\cite{quantum-fourier-transform-qiskit2022} to illustrate the execution time of the quantum circuits with different numbers of qubits.

\begin{figure}[htb]
\vspace{-0.4cm}
 \centering
 \captionsetup{justification=centering}
  \subfloat[The execution time of QFT.]{\label{fig:exetime-quantum-fourier-transform}\includegraphics[width=0.25\textwidth]{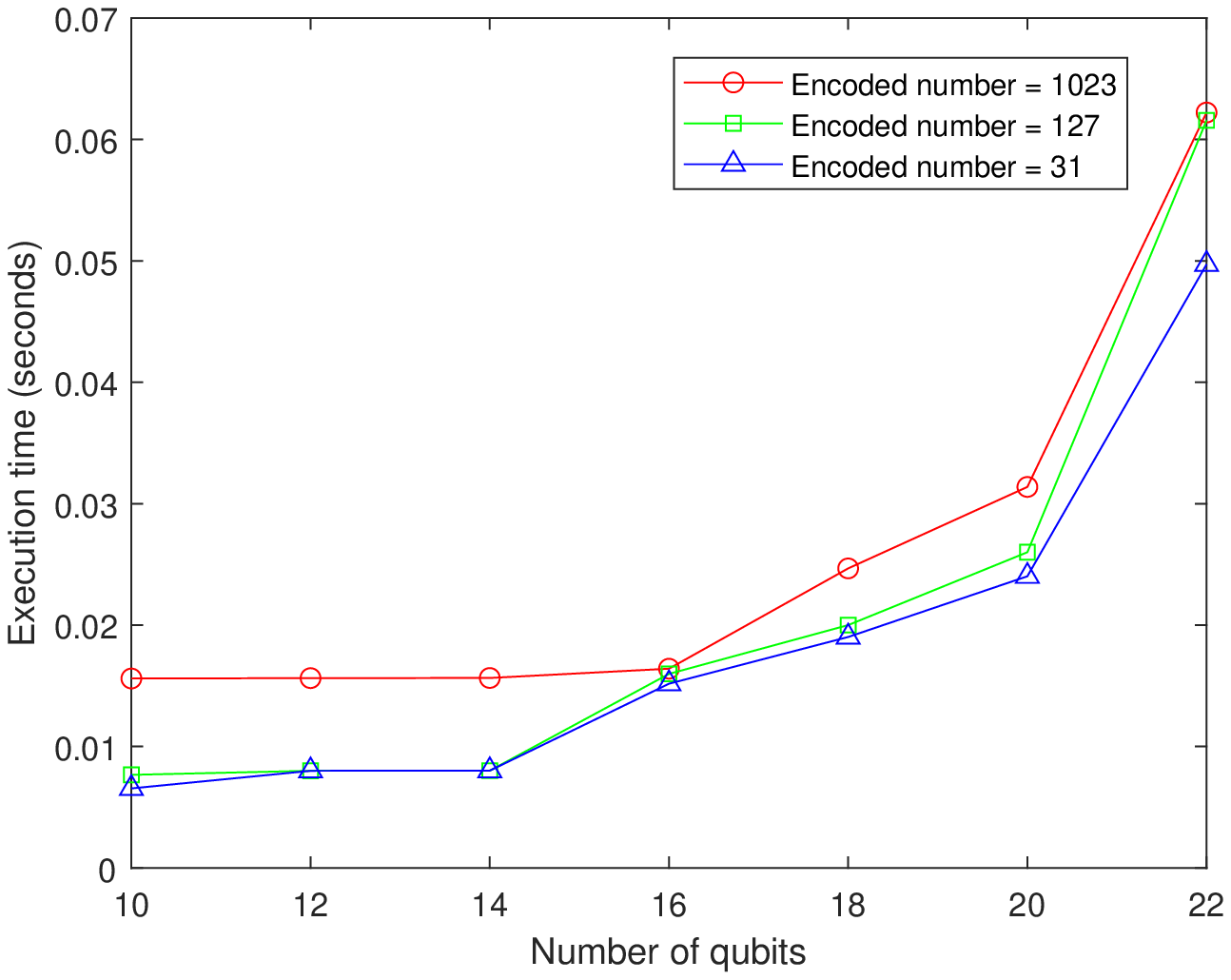}}
  \subfloat[The optimal solution under qubits.]{\label{fig:optimal-solution-varying-reserved-qubits}\includegraphics[width=0.25\textwidth]{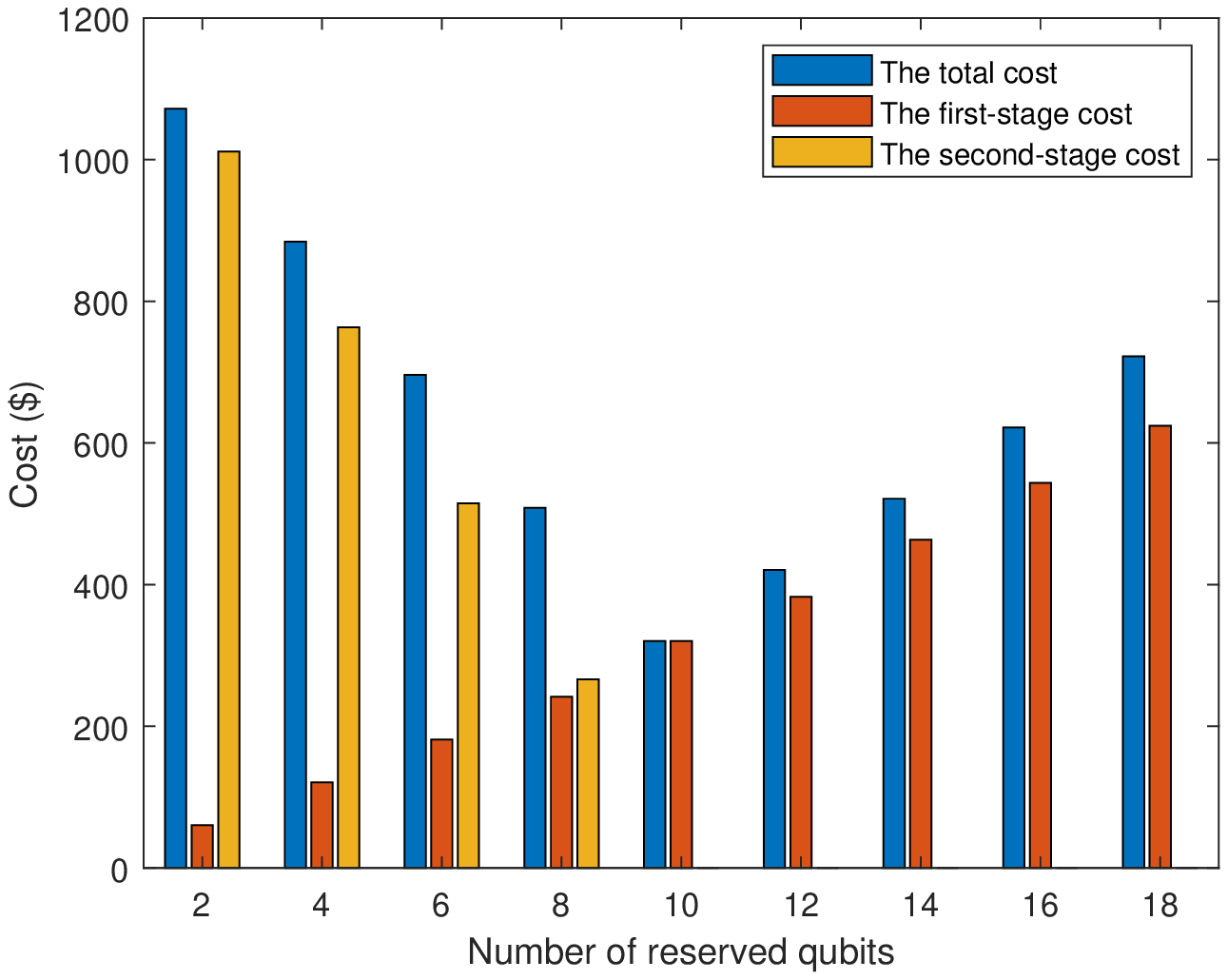}} \\ 
  \subfloat[The solution under qubits and waiting time.]{\label{fig:optimal-solution-varying-qubits-waiting-time}\includegraphics[width=0.35\textwidth]{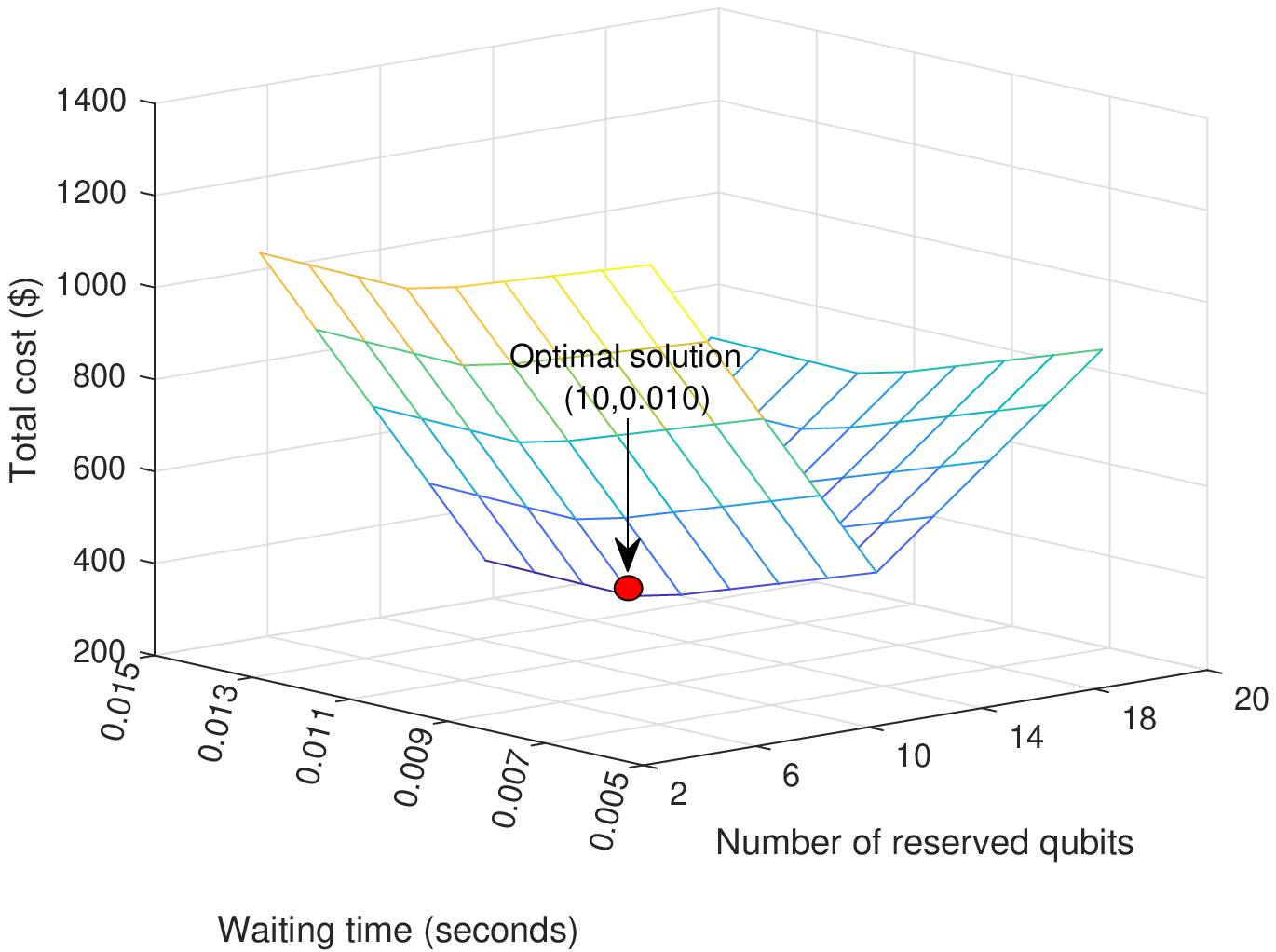}}  
 \caption{(a) The execution time of QFT under different encoded numbers ($\ket{\tilde{x}}$), (b) The optimal solution of SP model under different numbers of reserved qubits for QFT's quantum circuits, and (c) The solution of SP model under different numbers of reserved qubits and waiting time.}
 \label{fig:exe-time-solutions}
 \vspace{-0.4cm}
\end{figure}

Figure~\ref{fig:exe-time-solutions}(a) illustrates the execution time of QFT implemented by Qiskit~\cite{quantum-fourier-transform-qiskit2022} under different encoded numbers. In Fig.~\ref{fig:exe-time-solutions}(a), the execution time of QFT with the encoded number = 1023 (i.e., $\ket{\tilde{x}} = \ket{\tilde{1023}}$) is the highest value since the high number of qubits is used to represent the encoded number and to be calculated in QFT. For example, the number 1023 is represented by 10 qubits which are 1111111111 in binary representation and the depth of the quantum circuit of the encoded number = 1023 has the longest path. Therefore, we can explain that the high value of the encoded number and the depth of the quantum circuit directly affect the execution time that is obtained by the Qiskit simulation of the QFT.

\subsubsection{Cost Structure Analysis} 

We illustrate the performance of the SP model to achieve the optimal solution which is illustrated in Figs.~\ref{fig:exe-time-solutions}(b) and~\ref{fig:exe-time-solutions}(c). In addition, we present the effects of the number of over- and under-reserved qubits and the under-arranged waiting time of the quantum circuit on the solution. In Fig.~\ref{fig:exe-time-solutions}(b), we present the optimal solution in the case of the number of reserved qubits varied. In Fig.~\ref{fig:exe-time-solutions}(c), we present the optimal solution in the case of the number of reserved qubits and the arranged waiting time varied. 

In Fig.~\ref{fig:exe-time-solutions}(b), the first-stage cost linearly increases when the number of reserved qubits grows. However, the second-stage cost sharply decreases when the required number of qubits is observed in the second stage. This is due to the fact that the cost of qubits in the on-demand phase (i.e., the second stage) is higher than that of qubits in the reservation phase (i.e., the first stage). As a result, at the number of reserved qubits of 10, the optimal solution that is the first-stage cost is equal to the total cost obtained while the second-stage cost is 0. The reason is that the number of reserved qubits in the first stage satisfies the demands (i.e., the number of qubits required by quantum circuits) and therefore the number of on-demand qubits in the second stage does not utilize. In addition, after the number of reserved qubits is 10, the total cost and the first-stage cost still increase since it has a penalty cost to be charged for the number of over-reserved qubits. Therefore, over-provisioning and under-provisioning of the number of qubits have a significant effect on the total cost and the first-stage cost. In Fig.~\ref{fig:exe-time-solutions}(c), we investigate the optimal solution when the number of reserved qubits and the waiting time of quantum circuits are varied. The results are clearly shown that the optimal solution as illustrated in Fig.~\ref{fig:exe-time-solutions}(c) is obtained by the SP model. The optimal solution in Fig.~\ref{fig:exe-time-solutions}(c) is (10, 0.010) which means that the optimal number of reserved qubits is 10 qubits while the arranged waiting time is 0.010 seconds. In addition, we observe that the total cost in Fig.~\ref{fig:exe-time-solutions}(c) is higher than the total cost in Fig.~\ref{fig:exe-time-solutions}(b) since the penalty cost of under-arranged waiting time of the quantum circuits is charged to the total cost in Fig.~\ref{fig:exe-time-solutions}(c). The under-arranged waiting time of the quantum circuit is from the execution time of the quantum circuit executed on the quantum computer of the provider ($t^{\mathrm{exe}}_{c,p,m}$) minus the arranged waiting time of the quantum circuit ($\bar{\alpha}_{c,\omega}$).        

\section{Conclusion}
\label{sec:conclusion}

In this paper, we have proposed the qubit resource allocation for quantum circuits in the quantum cloud computing system, where the number of qubits and the minimum waiting time of quantum circuits are efficiently provisioned. We have formulated the qubit resource allocation for quantum circuits as the two-stage SP model to minimize the total costs of qubits and the waiting time under uncertainties of the qubit requirement and the waiting time. In experiments, practical quantum circuits of the QFT have been applied to evaluate the performance of the proposed model. The results have clearly shown that the proposed model has distinctly achieved the optimal costs of qubits and waiting time. In addition, you can look at~\cite{r-kaewpuang-stochastic-qubit2022} for more information of this paper.

For future work, we will investigate the impact of various probability distributions of the qubit requirement and the waiting time on the total cost. In addition, we will formulate the multi-stage SP model to solve optimization problems with many decision stages of qubit resource provision.

\section*{Acknowledgment} 
\begin{footnotesize}
This research is supported in part by the NSFC under grant No. 62102099, the National Research Foundation (NRF), Singapore and Infocomm Media Development Authority under the Future Communications Research Development Programme (FCP), and DSO National Laboratories under the AI Singapore Programme (AISG Award No: AISG2-RP-2020-019), under Energy Research Test-Bed and Industry Partnership Funding Initiative, part of the Energy Grid (EG) 2.0 programme, under DesCartes and the Campus for Research Excellence and Technological Enterprise (CREATE) programme, Alibaba Group through Alibaba Innovative Research (AIR) Program and Alibaba-NTU Singapore Joint Research Institute (JRI); and Nanyang Technological University, Nanyang Assistant Professorship. In addition, the research is supported by the National Research Foundation (NRF) and Infocomm Media Development Authority under the Future Communications Research Development Programme (FCP). The research is also supported by the SUTD SRG-ISTD-2021-165, the SUTD-ZJU IDEA Grant (SUTD-ZJU (VP) 202102), and the Ministry of Education, Singapore, under its SUTD Kickstarter Initiative (SKI 20210204).
\end{footnotesize}

\end{document}